\documentclass{iau} 
\usepackage{graphicx}

\title[JD 11.~~The uniqueness of astronomical observatory publications] 
{The uniqueness of astronomical \\ observatory publications}

\author[Ole Ellegaard \& Bertil F.\ Dorch]   
{Ole Ellegaard$^1$
 \and Bertil F. Dorch$^{1,2}$}

\affiliation{$^1$University Library of Southern Denmark, \\ SDU, Campusvej 55, 5230 Odense, Denmark \\ email: {\tt oleell@bib.sdu.dk} \\[\affilskip]
$^2$Department of Physics, Chemistry and Pharmacy,  \\SDU, Campusvej 55, 5230 Odense, Denmark}

\pubyear{2020}
\volume{367}  
\jname{Education and Heritage in the Era of Big Data in Astronomy}
\editors{R.M. Ros, B. Garcia, S. Gullberg, J. Moldon \& P. Rojo, eds.}
\begin{document}

\maketitle

\begin{abstract}
Astronomical observatory publications include the work of local astronomers from observatories around the world and are traditionally exchanged between observatories through their libraries. However, large collections of these publications appear to be rare and are often incomplete. In order to assess the unique properties of the collections, we compare observatories present in our own collection from the university at Copenhagen, Denmark with two collections from the USA: one at the Woodman Library at Wisconsin-Madison and another at the Dudley Observatory in Loudonville, New York.

\keywords{Observatory publications, Historical collections}
\end{abstract}

\firstsection 
\section{Introduction}

Astronomical observatory publications (AOPs) are considered to be of a high scientific standard and is guaranteed through the reputation of the individual observatory and in their modern form dates back to the middle of the 18th century \cite[(Holl \& Vargha 2003)]{HollVargha03}. Initially, AOPs present the results of the institutions’ own observations obtained with local equipment and issued with or without peer review. AOPs have been a cheap and very popular form of knowledge exchange, and for many low-budget observatories an indispensable source of information. Today, almost no observatories exchange physical material. Instead, information is published in international journal articles or exchanged either via institutional repositories or websites, but problems remain in terms of storage, registration and retrieval of the material. 
Older AOPs can still be relevant and valuable. An example is photometry of variable stars. Many AOPs are listed in NASA’s ADS database, and many documents have been scanned (cf. adsabs.harvard.edu/historical.html) but the coverage is far from complete. Large physical collections of AOPs are rare and e.g. special examples are found in the United States at the Woodman Library at Wisconsin-Madison (W-M) (cf. www.library.wisc.edu/astronomy) and the Dudley Observatory (Dud)
(cf. dudleyobservatory.org/collections-overview).
The library at the University of Southern Denmark (SDU) has recently acquired a large collection of observatory publications from the now discontinued, historical library collections at the Niels Bohr Institute, University of Copenhagen. The collection’s material dates from a period of several centuries and consists of tables of observations, annual reports, bulletins, circulars, newsletters, reprints etc. It has been collected for a period of more than a hundred years at the observatory in Copenhagen. The oldest material in the collection dates back to the 1700s. To understand the importance and coverage, we examine the difference between the material found in the various collections. In practice, we compare the number of cities/observatories present.

\section{Results}

Publications of all three library collections are registered under the city where the observatory is located. For some of the largest cities are more than one observatory registered.

\begin{figure}[h]
\begin{center}
 \includegraphics[width=5.3in]{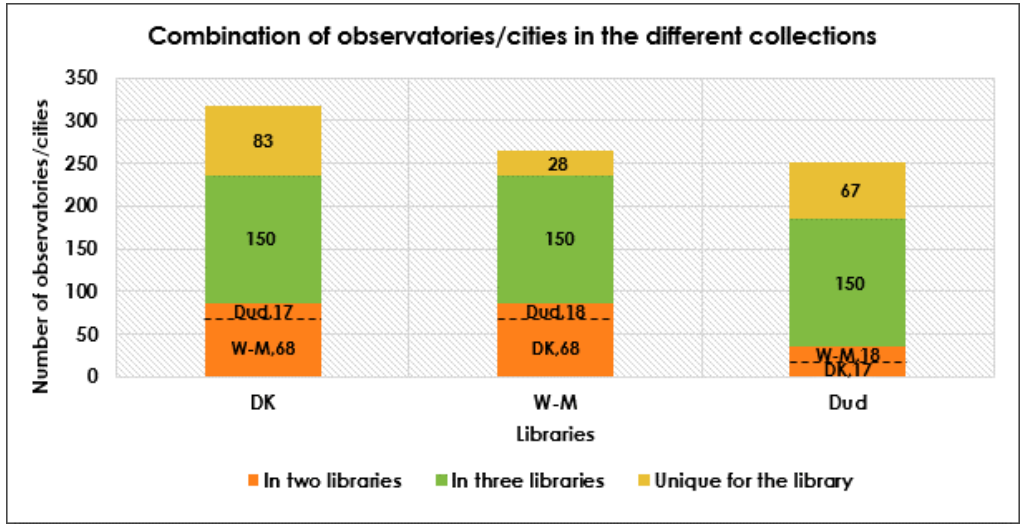}

 \caption{Combination of cities/observatories in the various collections.}
   \label{fig1}
\end{center}
\end{figure}

318 observatories are found in the danish collection, 264 in the W-M collection and 252 in the Dudley collection (Fig.\,\ref{fig1}). There is a overlap between the observatories present in the three collections. 83 or (26\%) are unique and can only be found in the Danish collection, 28 (11\%) in the W-M collection and 67 (27\%) in the Dudley collection. The Dudley collection is the smallest but with the relatively largest number of unique observatories.

\section{Discussion}

The traditional AOP has now been replaced by the more widespread publication in traditional journals. A few libraries around the world still have quite large collections of older AOPs in paper format. Some of this material is included in the ADS, and is based on the collection at the astronomical institution at Wolbach Library, Harvard and supplemented by other collections, but other publications can only be found by searching the more or less complete registers of the relevant libraries. For example, the register in the Danish collection is present as a card catalog, and the Dudley collection is registered as incomplete items that are transferred directly from the cards. Not all cities/observatories are represented in the individual collections, as shown by the current study. The Danish collection represents the largest number of unique cities/observatories, followed by the Dudley Collection. We encourage astronomers looking for different types of older, rare materials that are hard to find, to be aware of and to search in ADS or in at least one major library that has a large collection of observatory publications.

{}


\begin{thebibliography}{}


\bibitem[Holl \& Vargha (2003)]{HollVargha03}
Holl. A., \& Vargha, M. 2003, In: B. Corbin, E. Bryson, \&  M. Wolf (eds.), \textit{ Observatory Publications—Quo Vadis?},  Library and Information Services in Astronomy IV (LISA IV), Prague, Czech Republic, p.109-116






\end{thebibliography}
\end{document}